\documentclass[a4paper,10pt,twoside]{cpc-hepnp}
\usepackage{CJK,upgreek,fancyhdr}
\usepackage{multicol}
\usepackage{graphicx}
\usepackage{booktabs}
\usepackage{amssymb,bm,mathrsfs,bbm,amscd}
\usepackage[tbtags]{amsmath}
\usepackage{lastpage}
\usepackage[english]{babel}
\usepackage{epsfig}
\usepackage{graphicx}
\usepackage{epstopdf}
\usepackage{float}

\begin{document}
\begin{CJK*}{GBK}{song}

\fancyhead[c]{\small Chinese Physics C~~~Vol. xx, No. x (2022) xxxxxx}
\fancyfoot[C]{\small 010201-\thepage}
\footnotetext[0]{Received \today}

\title{Deep learning method in testing the cosmic distance duality relation\thanks{Supported by the National Natural Science Fund of China (Grant Nos. 11873001 and 12147102), and the Fundamental Research Funds for the Central Universities of China (Grants No. 2022CDJXY-002)}}

\author{%
Li Tang$^{1;2}$
\quad Hai-Nan Lin$^{2;3;1)}$\email{linhn@cqu.edu.cn (Corresponding author)}
\quad Liang Liu$^{1;2}$
}
\maketitle

\address{%
$^1$ Department of Math and Physics, Mianyang Normal University, Mianyang 621000, China\\
$^2$ Department of Physics, Chongqing University, Chongqing 401331, China\\
$^3$ Chongqing Key Laboratory for Strongly Coupled Physics, Chongqing University, Chongqing 401331, China\\
}

\begin{abstract}
  The cosmic distance duality relation (DDR) is constrained from the combination of type-Ia supernovae (SNe Ia) and strong gravitational lensing (SGL) systems using deep learning method. To make use of the full SGL data, we reconstruct the luminosity distance from SNe Ia up to the highest redshift of SGL using deep learning, then it is compared with the angular diameter distance obtained from SGL. Considering the influence of lens mass profile, we constrain the possible violation of DDR in three lens mass models. Results show that in the SIS model and EPL model, DDR is violated at high confidence level, with the violation parameter $\eta_0=-0.193^{+0.021}_{-0.019}$ and $\eta_0=-0.247^{+0.014}_{-0.013}$, respectively. In the PL model, however, DDR is verified within 1$\sigma$ confidence level, with the violation parameter $\eta_0=-0.014^{+0.053}_{-0.045}$. Our results demonstrate that the constraints on DDR strongly depend on the lens mass models. Given a specific lens mass model, DDR can be constrained at a precision of $\textit{O}(10^{-2})$ using deep learning.
\end{abstract}



\footnotetext[0]{\hspace*{-3mm}\raisebox{0.3ex}{$\scriptstyle\copyright$}2019
Chinese Physical Society and the Institute of High Energy Physics
of the Chinese Academy of Sciences and the Institute
of Modern Physics of the Chinese Academy of Sciences and IOP Publishing Ltd}%

\begin{multicols}{2}

\section{Introduction}\label{sec:introduction}

In the expanding Universe there are many ways to define the distance between two objects, among which the luminosity distance and angular diameter distance are two widely used definitions. The former is defined by the fact that the brightness of a distant source seen by an observer is inversely proportion to the distance squared, while the latter is defined by the fact that the angular size of an object seen by an observer is inversely proportion to the distance. In the standard cosmological model, the luminosity distance ($D_L$) is related to the angular diameter distance ($D_A$) by the distance duality relation (DDR), i.e. $D_L(z)=(1+z)^2D_A(z)$ \cite{Etherington1933}, where $z$ is the cosmic redshift. DDR holds true as long as the photons travel along null geodesics and the number of photons is conserved. DDR is a fundamental and crucial relation in modern cosmology. Any violation of DDR would imply the existence of new physics. The violation of DDR could be caused by e.g. the coupling of photons with non-standard particles \cite{Bassett2004}, the dust extinction \cite{Corasaniti2006}, the variation of fundamental constants \cite{Eills2013}, etc. Therefore, testing the validity of DDR with different independent observations is of great importance.

Due to the difficulty to measure the luminosity distance and angular diameter distance of an object simultaneously, the most common way to test DDR is comparing the two distances observed from different objects but approximately at the same redshift. Many works have been devoted to testing DDR with different observational data \cite{Holanda_2010,Li_2011,Liang2013,Liao_2016,Li:2017zrx,Lin:2019mrl,Lin:2020vqj,Arjona:2020axn,Lima:2021slf}. Type Ia supernovae (SNe Ia) are perfect standard candles and are widely used to measure the luminosity distance. While the angular diameter distance is usually obtained from various observations. For example, the angular diameter distance obtained from galaxy clusters can be combined with SNe Ia to test DDR \cite{Holanda_2010,Li_2011,Liang2013,Lima:2021slf}. Li et al. \cite{Li:2017zrx} tested DDR using the combination of SNe Ia and ultra-compact radio sources. Liao et al. \cite{Liao_2016} proposed a model-independent method applying strong gravitational lensing (SGL) systems and SNe Ia to test DDR. However, due to the redshift limitation of SNe Ia, SGL systems whose source redshift is larger than 1.4 couldn't be used to test DDR, since no SNe can match these SGL system at the same redshift. Hence, the available data pairs are much less than the total number of SGL systems. Lin et al. \cite{Lin:2019mrl,Lin:2020vqj} shown that the luminosity distance and angular diameter distance can be measured simultaneously from the strongly lensed gravitational waves, hence can be used to test DDR.

Recently, a newest and largest SNe Ia sample called Pantheon compilation was published, which consists of 1048 SNe Ia and the highest redshift is up to $z_{\rm max}\approx 2.3$ \cite{Scolnic:2018}. Combining the Pantheon sample and 205 SGL systems, Zhou et al. \cite{Zhou_2021} obtained 120 pairs of data points in the redshifts range from 0.11 to 2.16 and verified DDR at 1$\sigma$ confidence level. Many methods are proposed to extend the redshift range. Lin et al. \cite{Lin:2018qal} reconstructed the distance-redshift relation from the Pantheon sample with Gaussian processes (GP), and combined it with the galaxy clusters + baryon acoustic oscillations to constrain DDR. Their results verified the validity of DDR. Ruan et al. \cite{Ruan:2018dls} also confirmed the validity of DDR in a redshift range up to $z\sim2.33$ based on strong gravitational lensing and the reconstruction of HII galaxy Hubble diagram using GP. However, the reconstruction with GP is unreliable beyond the data region, and the uncertainty is very large in the region where the data points are sparse. Hence not all SGL systems are available to test DDR. To make use of the full SGL sample, Qin et al. \cite{Qin:2021jqy} reconstructed the high-redshift quasar Hubble diagram using the B$\acute{\rm e}$zier parametric fit, and combined it with 161 SGL systems to test DDR, up to redshift $z\sim3.6$. However, this method depends on the parametric form of the Hubble diagram. Some works extended the luminosity distance to the redshift range of gamma-ray burst (GRBs) \cite{Holanda:2016msr,Fu:2017nmw}. Combined SNe Ia and high-redshift GRBs data ($z\sim10$) calibrated with the Amati relation, DDR can be investigated up to high redshift. In these works, the Amati correlation used to calibrate GRBs is assumed to be universal over full redshift range. However, several works indicate that the Amati relation possibly evolves with redshift \cite{Li:2007mt,Lin:2015fma,Tang:2020nmw}.

In this paper, we test DDR by performing the deep learning method to reconstruct the luminosity distance. The deep learning is one kind of machine learning based on the artificial neural network research \cite{Aurelien:2017}. It is powerful in large-scale data processing and highly complex data fitting due to the universal approximation theorem. Hence, deep learning can output a value infinitely approximating the target by training deep neural networks with the observational data. Generally, a neural network is constructed with several layers. Thereinto, the first layer is input layer for receiving the feature, several hidden layers for transforming the information from previous layer, and the last layer is output layer for exporting the target. In each layer, hundreds of nonlinear neurons process the data information. The deep learning method has been widely employed in various fields including cosmological research \cite{Escamilla-Rivera:2019hqt,Wang:2019vxv,Tang:2021elg,Tang:2020nmw,Arjona:2020axn,Liu:2021fka}. In this work, we test DDR with the combined data of SNe Ia and SGL, while the luminosity distance is reconstructed from SNe Ia using deep learning. Compared with previous methods such as GP and B$\acute{\rm e}$zier parametric fit, our method to reconstruct the data is neither cosmologically model-dependent, nor relies on the specific parametric form. Additionally, deep learning can captures the internal relation in training data. Especially, deep learning can reconstruct the curve beyond the data range with a relatively small uncertainty. Thus the full SGL sample can be used to test DDR, up to the highest redshift of SGL.

The rest of this paper is arranged as follows: In Section 2, we introduce the method to test DDR with the combination of SGL and SNe Ia. In Section 3, the observational data and deep learning method are introduced. In Section 4, the constraining results on DDR are presented. Finally, discussion and conclusion are given in Section 5.

\section{Methodology}\label{sec:methodology}

The most direct way to test DDR is to compare the luminosity distance $D_L$ and angular diameter distance $D_A$ at the same redshift. However, it is difficult to measure $D_L$ and $D_A$ simultaneously from a single object. Generally, $D_L$ and $D_A$ are obtained from different kinds of objects approximately at the same redshift. In our work, we determine $D_L$ from SNe Ia and determine $D_A$ from SGL systems, respectively. We present the details of the method  below.

As more and more galaxy-scale SGL systems are discovered in recent years, they are widely used to investigate the gravity and cosmology. In the specific case when the lens perfectly aligns with the source and the observer, an Einstein ring appears. In general case only a part of the ring appears, from which the radius of the Einstein ring can be deduced. The Einstein radius not only depends on the geometry of the lensing system, but also depends on the mass profile of the lens galaxy. To investigate the influence of mass profile of lens galaxy, we consider three types of lens models that are widely discussed in literature, i.e. the singular isothermal sphere (SIS) model, the power-law (PL) model and the extended power-law (EPL) model.

In the singular isothermal sphere (SIS) model, the mass density of the lens galaxy scales as $\rho\propto r^{-2}$, and the Einstein radius takes the form \cite{Mollerachc:2002}
\begin{equation}\label{eq:theta_E}
\theta_{\rm E}=\frac{D_{A_{ls}}}{D_{A_s}}\frac{4\pi\sigma^2_{\rm{SIS}}}{c^2},
\end{equation}
where $c$ is the speed of light in vacuum, $\sigma_{\rm SIS}$ is the velocity dispersion of the lens galaxy, $D_{A_s}$ and $D_{A_{ls}}$ are the angular diameter distances between the observer and the source, and between the lens and the source, respectively. From Eq.(\ref{eq:theta_E}) we see that the Einstein radius depends on the distance ratio $R_A\equiv D_{A_{ls}}/D_{A_s}$, which can be obtained from the observables $\theta_{\rm E}$ and $\sigma_{\rm SIS}$ by
\begin{equation}\label{eq:SIE}
R_A=\frac{c^2\theta_{\rm{E}}}{4\pi\sigma^2_{\rm{SIS}}}.
\end{equation}
Note that $\sigma_{\rm{SIS}}$ is not necessary to be equal to the observed stellar velocity dispersion $\sigma_0$ \cite{Khedekar2011}. Thus we phenomenological introduce a parameter $f$ to account for the difference, i.e. $\sigma_{\rm{SIS}}=f\sigma_0$ \cite{Kochanek1992,Ofek:2003sp,Cao:2012}. Here, $f$ is a free parameter which is expected to be in the range $0.8<f^2<1.2$. The actual SGL data usually measure the velocity dispersion within the aperture radius $\theta_{\rm ap}$, which can be transformed to $\sigma_0$ according to the aperture correction formula \cite{Jorgensen:1995zz}
\begin{equation}
\sigma_0=\sigma_{\rm ap}\left(\frac{\theta_{\rm{eff}}}{2\theta_{\rm{ap}}}\right)^{\eta},
\end{equation}
where $\sigma_{\rm ap}$ is the luminosity weighted average of the line-of-sight velocity dispersion inside the aperture radius, $\theta_{\rm{eff}}$ is the effective angular radius, and $\eta$ is the correction factor which takes the value $\eta=-0.066\pm0.035$ \cite{Cappellari:2005ux,Chen:2018jcf}. $\sigma_{\rm ap}$ propagates its uncertainty to $\sigma_0$, and further to $\sigma_{\rm SIS}$. The uncertainty of distance ratio $R_A$ is propagated from that of $\theta_E$ and $\sigma_{\rm SIS}$. We take the fractional uncertainty of $\theta_E$ at the level of $5\%$ \cite{Liao_2016}.

In the power-law (PL) model, the mass density of the lensing galaxy follows the spherically symmetric power-law distribution $\rho\propto r^{-\gamma}$, where $\gamma$ is the power-law index. In this model, the distance ratio can be written as \cite{Koopmans:2006iu}
\begin{equation}\label{eq:PLS}
R_A=\frac{c^2\theta_{\rm{E}}}{4\pi\sigma^2_{\rm{ap}}}\left(\frac{\theta_{\rm ap}}{\theta_{\rm E}}\right)^{2-\gamma}f^{-1}(\gamma),
\end{equation}
where
\begin{equation}
f(\gamma)=-\frac{1}{\sqrt{\pi}}\frac{(5-2\gamma)(1-\gamma)}{3-\gamma}\frac{\Gamma(\gamma-1)}{\Gamma(\gamma-3/2)}\left[\frac{\Gamma(\gamma/2-1/2)}{\Gamma(\gamma/2)}\right]^2.
\end{equation}
The power-law model reduces to the SIS model when $\gamma=2$. Considering the possible redshift evolution of the mass density profile, we parameterize $\gamma$ with the form $\gamma(z_l)=\gamma_0+\gamma_1z_l$, where $\gamma_0$ and $\gamma_1$ are two free parameters, and $z_l$ is the redshift of the lens galaxy.

In the extended power-law (EPL) model, the luminosity density profile $\nu(r)$ is usually different from the total-mass density profile $\rho(r)$ due to the contribution of dark matter halo. Therefore, we assume that the total mass density profile $\rho(r)$ and the luminosity density of stars $\nu(r)$ respectively take the forms as
\begin{equation}
\rho(r)=\rho_0\left(\frac{r}{r_0}\right)^{-\alpha}, \ \nu(r)=\nu_0\left(\frac{r}{r_0}\right)^{-\delta},
\end{equation}
where $\alpha$ and $\delta$ are the power-law index parameters, $r_0$ is the characteristic length scale, $\rho_0$ and $\nu_0$ are two normalization constants. In this case, the distance ratio can be expressed as \cite{Birrer:2018vtm}
\begin{equation}\label{eq:EPL}
R_A=\frac{c^2\theta_{\rm{E}}}{2\sigma_0^2\sqrt{\pi}}\frac{3-\delta}{(\xi-2\beta)(3-\xi)}\left(\frac{\theta_{\rm eff}}{\theta_E}\right)^{2-\alpha}\left[\frac{\lambda(\xi)-\beta\lambda(\xi+2)}{\lambda(\alpha)\lambda(\delta)}\right],
\end{equation}
where $\xi=\alpha+\delta-2$, $\lambda(x)=\Gamma(\frac{x-1}{2})/\Gamma(\frac{x}{2})$, and $\beta$ is an anisotropy parameter characterizing the anisotropic distribution of the three-dimensional velocity dispersion, which is marginalized with Gaussian prior $\beta=0.18\pm0.13$ \cite{Wang:2019yob}. We parameterize $\alpha$ with the form $\alpha=\alpha_0+\alpha_1z_l$ to inspect the possible redshift-dependence of the lens mass profile, and treat $\delta$ as a free parameter. When $\alpha_0=\delta=2$ and $\alpha_1=\beta=0$, the EPL model reduces to the standard SIS model.

In the flat Universe, the comoving distance is related to the angular diameter distance by $r_l=(1+z_l)D_{A_l}$, $r_s=(1+z_s)D_{A_s}$, $r_{ls}=(1+z_s)D_{A_{ls}}$. Using the distance-sum rule $r_{ls}=r_s-r_l$ \cite{Rasanen:2014mca}, the distance ratio $R_A$ can be expressed as
\begin{equation}\label{eq:RA2}
R_A=\frac{D_{A_{ls}}}{D_{A_s}}=1-\frac{(1+z_l)D_{A_l}}{(1+z_s)D_{A_s}},
\end{equation}
in which the ratio of $D_{A_l}$ and $D_{A_s}$ can be converted to the ratio of $D_{L_l}$ and $D_{L_s}$ using DDR.

To test the possible violation of DDR, we parameterize it with the form
\begin{equation}\label{eq:eta0}
\frac{D_A(z)(1+z)^2}{D_L(z)}=1+\eta_0z,
\end{equation}
where $\eta_0$ is a parameter characterizing the deviation from DDR. The standard DDR is the case when $\eta_0\equiv0$. Combining Eqs.(\ref{eq:RA2}) and (\ref{eq:eta0}), we obtain
\begin{equation}\label{eq:R_A_R_L}
R_A(z_l,z_s)=1-R_L P(\eta_0;z_l,z_s),
\end{equation}
where $R_L\equiv D_{L_l}/D_{L_s}$ is the ratio of luminosity distance, and
\begin{equation}
P\equiv\frac{(1+z_s)(1+\eta_0z_l)}{(1+z_l)(1+\eta_0z_s)}.
\end{equation}

The ratio of luminosity distance $R_L$ can be obtained from SNe Ia. At a certain redshift $z$, the distance modulus of SNe Ia is given by \cite{Scolnic:2018}
\begin{equation}\label{eq:muDL}
\mu=5\log_{10}\frac{D_L(z)}{\textrm{Mpc}}+25=m_B-M_B+\alpha x(z)-\beta c(z),
\end{equation}
where $m_B$ is the apparent magnitude observed in B-band, $M_B$ is the absolute magnitude, $x$ and $c$ are the stretch and colour parameters respectively, $\alpha$ and $\beta$ are nuisance parameters. For SNe Ia sample, we choose the largest and latest Pantheon dataset in redshift range $z\in[0.01,2.30]$ \cite{Scolnic:2018}. The Pantheon sample is well-calibrated by a new method called BEAMS with Bias Corrections, and the effects of $x(z)$ and $c(z)$ have been corrected in the reported magnitude $m_{B,\textrm{corr}}=m_B+\alpha x(z)-\beta c(z)$. Thus, the nuisance parameters $\alpha$ and $\beta$ are fixed to zero in equation (\ref{eq:muDL}) and $m_B$ is replaced by the corrected magnitude $m_{B, \rm corr}$. For simplify, we use $m$ to represent $m_{B, \rm corr}$ hereafter.
Then the distance ratio $R_L$ can be written as
\begin{equation}\label{eq:R_L}
R_L\equiv D_{L_l}/D_{L_s}=10^{\frac{m(z_l)-m(z_s)}{5}},
\end{equation}
where $m(z_l)$ and $m(z_s)$ are the corrected apparent magnitudes of SNe Ia at redshifts $z_l$ and $z_s$, respectively. As is shown in the above equation, the absolute magnitude $M_B$ exactly cancels out. The uncertainty of $R_L$ propagates from the uncertainties of $m(z_l)$ and $m(z_s)$ using the standard error propagation formula.

Combining the SNe Ia and SGL, the parameter $\eta$ can be constrained by maximizing the likelihood
\begin{equation}\label{eq:L}
\mathcal{L}({\rm Data}|p,\eta_0)\propto\exp\left[-\frac{1}{2}\sum_{i=1}^N\frac{\left(1-R_{L,i}P_i(\eta_0;z_l,z_s)-R_{A,i}\right)^2}{\sigma_{{\rm total},i}^2}\right],
\end{equation}
where $\sigma_{\rm total}=\sqrt{\sigma_{R_A}^2+P^2\sigma_{R_L}^2}$ is the total uncertainty, and $N$ is the total number of data points. There are two sets of parameters, that is, the parameter $\eta_0$ relating to the violation of DDR, and the parameter $p$ relating to the lens mass model ($p=f$ in SIS model, $p=(\gamma_0,\gamma_1)$ in PL model and $p=(\alpha_0,\alpha_1,\delta)$ in EPL model).

\section{Deep Learning}\label{sec:deeplearing}

The SGL sample used in our paper are compiled in Chen et al. \cite{Chen:2018jcf}, which contains 161 galaxy-scale SGL systems with both high resolution imaging and stellar dynamical data. All lens galaxies in the SGL sample are early-type galaxies and don't have significant substructure or close massive companion. Thus the spherically symmetric approximation is valid when modelling the lens galaxy. The redshift of the lens ranges from 0.064 to 1.004, and that of the source ranges from 0.197 to 3.595. Hence, we can test DDR up to $z\sim3.6$.

In previous works \cite{Holanda_2010,Li_2011,Liao_2016}, SNe Ia locating at the redshift $z_l$ and $z_s$ in each SGL system are found by comparing the redshift difference $\Delta z$ between the lens (source) and SNe Ia with the criterion $\Delta z\leq0.005$. In this method, the SGL systems are under-utilized due to that there may be no SNe at redshift $z_l$ or $z_s$. To increase the available SGL sample, some works employed the GP method to reconstruct the distance-redshift relation from SNe Ia \cite{Ruan:2018dls,Wang:2019yob}. However, the SGL systems whose source redshift is higher than the maximum redshift of SNe is still unusable. Because the GP method cannot precisely reconstruct the curve beyond the redshift range of observational data. In order to make use of all SGL systems and match the lens and source redshifts of SGL sample with SNe Ia one-to-one, we apply a model-independent deep learning method to reconstruct the distance-redshift relation from SNe Ia, covering the full redshift range of SGL sample.

Deep learning is a dramatic method in discovering the intricate structures in a large and complex dataset by considering an Artificial Neural Networks (ANN) as an underlying model, such as Convolutional Neural Networks (CNN), Recurrent Neural Networks (RNN), Bayesian Neural Networks (BNN), etc. These neural networks are usually composed of multiple processing layers, in which each layer receives the information from the previous layer and transforms it to the next layer, and are trained to be an ideal network to represent the data. Thereinto, RNN is powerful in tackling the sequential data and predicting the future after learning the representation of the data. Hence, we can feed RNN with the Pantheon data to learn the distance-redshift relation, and train it to predict the distance at any redshift, even beyond the redshift range of the observational data. However, RNN is incapable to estimate the uncertainty of the prediction. Thus, we need to introduce BNN into our network as a supplementary of RNN to calculate the uncertainty of the prediction. In our recent paper \cite{Tang:2020nmw}, we have constructed a network combining RNN and BNN to represent the relationship between distance modulus $\mu$ and redshift $z$ from the Pantheon data. In this work, instead of reconstructing $\mu(z)$ curve, we reconstruct the apparent magnitude curve $m(z)$ from the Pantheon data, as the latter depends on neither the absolute magnitude of SNe nor the Hubble constant. Considering a constant difference between the distance moduli $\mu$ and the apparent magnitude $m$, we directly use our previous network to reconstruct $m$ but without setting the absolute magnitude and Hubble constant. The construction of our network is briefly introduced as follows (see Tang et al. \cite{Tang:2020nmw} for more details).

\begin{figure}[H]
  \centering
  \includegraphics[width=0.8\textwidth]{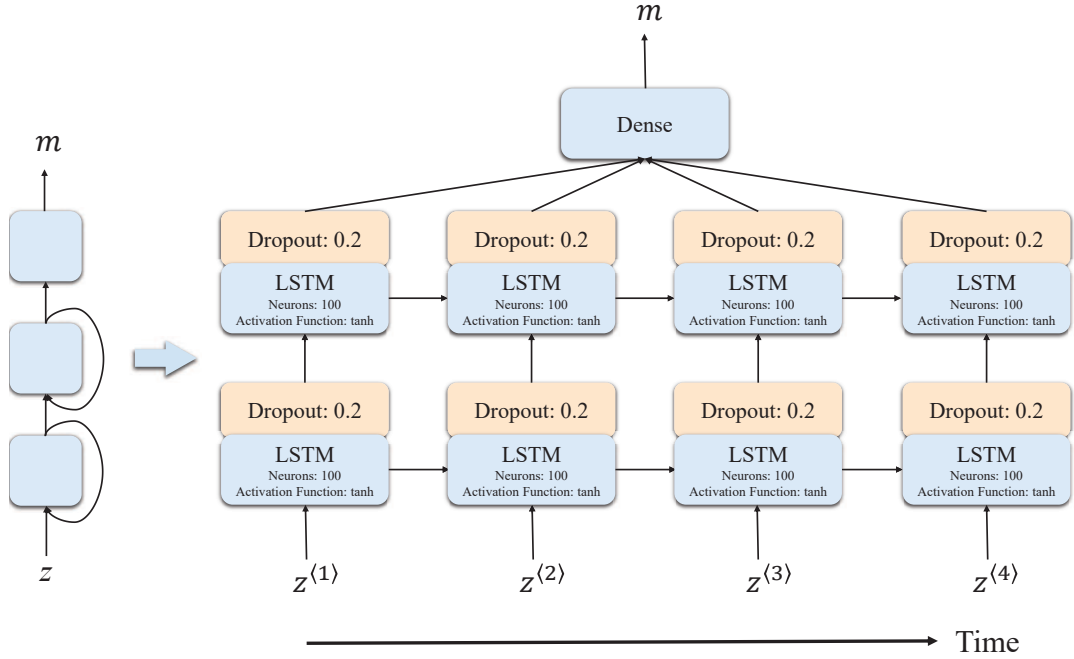}
  \caption{\small{The architecture of RNN with one hidden layer (left), unrolled through time step $t=4$ (right). The input and output are the redshift sequence and the corresponding apparent magnitudes, respectively. The number of neurons in each LSTM cell is 100. The activation function in each LSTM cell is tanh, and the dropout between two adjacent LSTM layers is set to 0.2.}}\label{fig_RNN}
\end{figure}

The architecture of our network is shown in Figure \ref{fig_RNN}. The main structure of RNN is composed of three layers, the input layer to receive the feature (the redshift $z$ here), one hidden layer to transform the information from the previous layer to the next layer, and the output layer to export the target (the apparent magnitude $m$ here). The information not only propagates forward from the first layer, through the hidden layer to the last layer, but also propagates backward. This can be seen more obviously from the unrolled RNN in the right panel of Figure \ref{fig_RNN}. At each time step $t$, the neurons of RNN receive the input as well as the output from the previous time step $t-1$. The RNN unfolded in time step can be regarded as a deep network where the number of hidden layers is more than 1. It takes long time to train RNN when handling long sequential data. Besides, RNN is difficult to store the information for long time. To solve this problem, we set the time step with $t=4$ and employ the Long Short-Term Memory (LSTM) cell as the basic cell. The LSTM cells augment the RNN with an explicit memory so that the network is aware of what to store, throw away and read. We built the input and hidden layers with LSTM cells of 100 neurons in each. Feeding in the Pantheon data, the RNN is trained to represent the relationship between magnitude $m$ and redshift $z$ by minimizing the loss function, which depicts the difference between the predictions and the observations. In our network, we choose the mean-squared-error (MSE) function as the loss function, and the Adam optimizer is adopted to find its minimum. Additionally, a non-linear activation function $A_f$ is introduced to enhance the performance of the network. In our previous work \cite{Tang:2020nmw}, we have shown that the tanh function \cite{karlik2011performance} performs better than other three activation functions (relu \cite{Abien2018}, elu \cite{Clevert2016} and selu \cite{Klambauer2017}) in reconstructing the distance moduli $\mu(z)$. Considering that the apparent magnitude is connected with the distance modulus with a overall constant, we directly choose the tanh function as the activation function.

\end{multicols}
\begin{multicols}{2}

For BNN, it should be mentioned that a traditional BNN is too complex to design. According to  Gal \& Ghahramani \cite{Gal:2016a,Gal:2016b,Gal:2016c}, the dropout in deep neural network can be seen as an approximate Bayesian inference in deep Gaussian processes. The dropout contributes an additional loss to the training process besides the difference between the predictions and the observations. Minimizing the objective relating to the total loss in the network results in the same optimal parameters as maximizing the objective log Evidence Lower Bound in Gaussian process approximation \cite{Gal:2016b}. In other words, network with a dropout is mathematically equivalent to the Bayesian model. Hence, we apply the dropout technique to RNN to realize BNN. When the RNN is well trained, the network executed $n$ times can determine the confidence region of the prediction, which is equivalent to BNN. Besides, the dropout is one kind of regularization techniques to prevent the network from over-fitting caused by a large number of its internal hyperparameters. Considering the risks of over-fitting and under-fitting, we set the dropout employed between two adjacent LSTM layers to 0.2 in this paper \cite{Muthukrishna:2019wgc,Bonjean:2019eux,Mangena:2020jdo}. The hyper-parameters used in our network are presented in Figure \ref{fig_RNN}.

\begin{figure}[htbp]
\centering
  \includegraphics[width=0.48\textwidth]{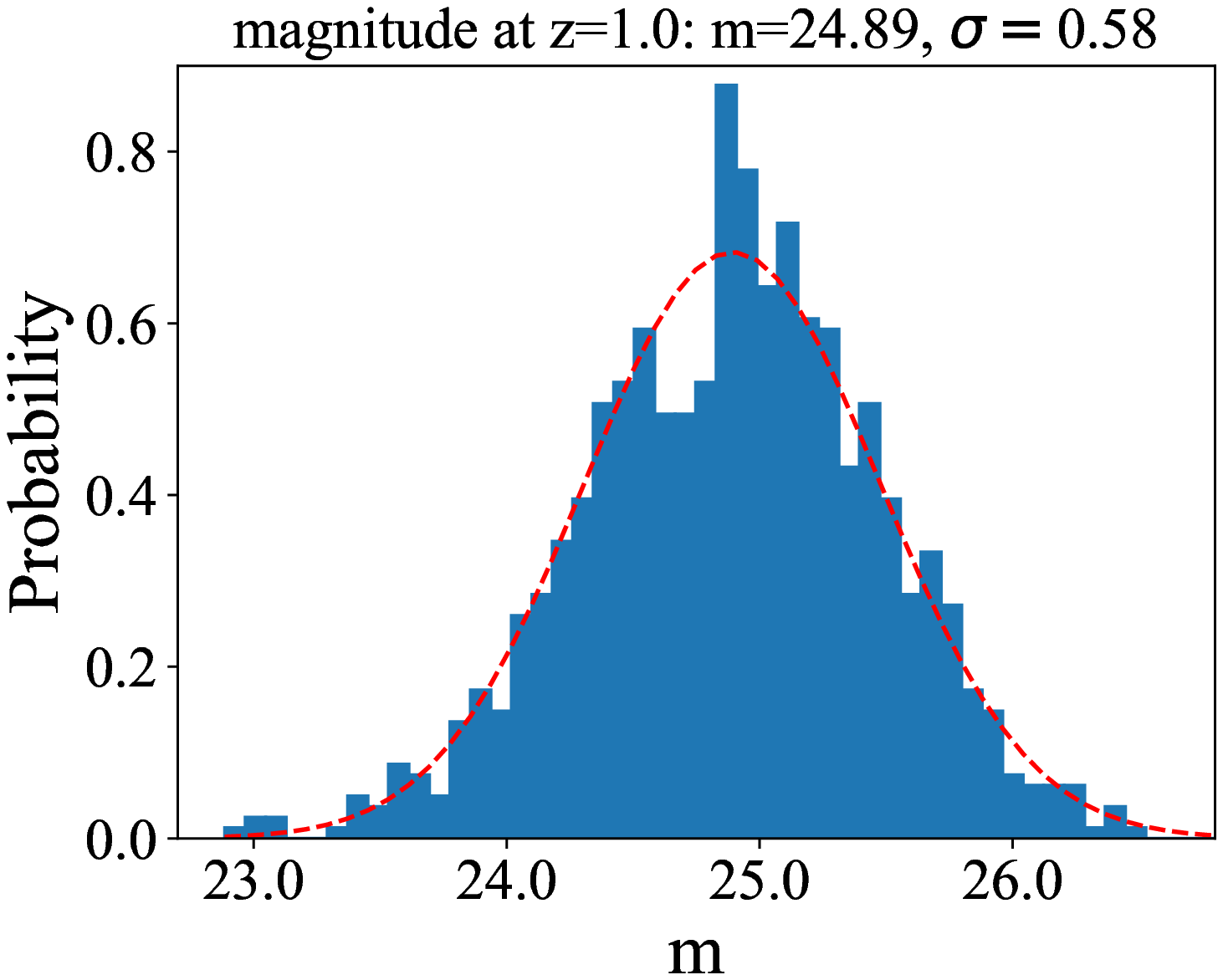}
  \includegraphics[width=0.48\textwidth]{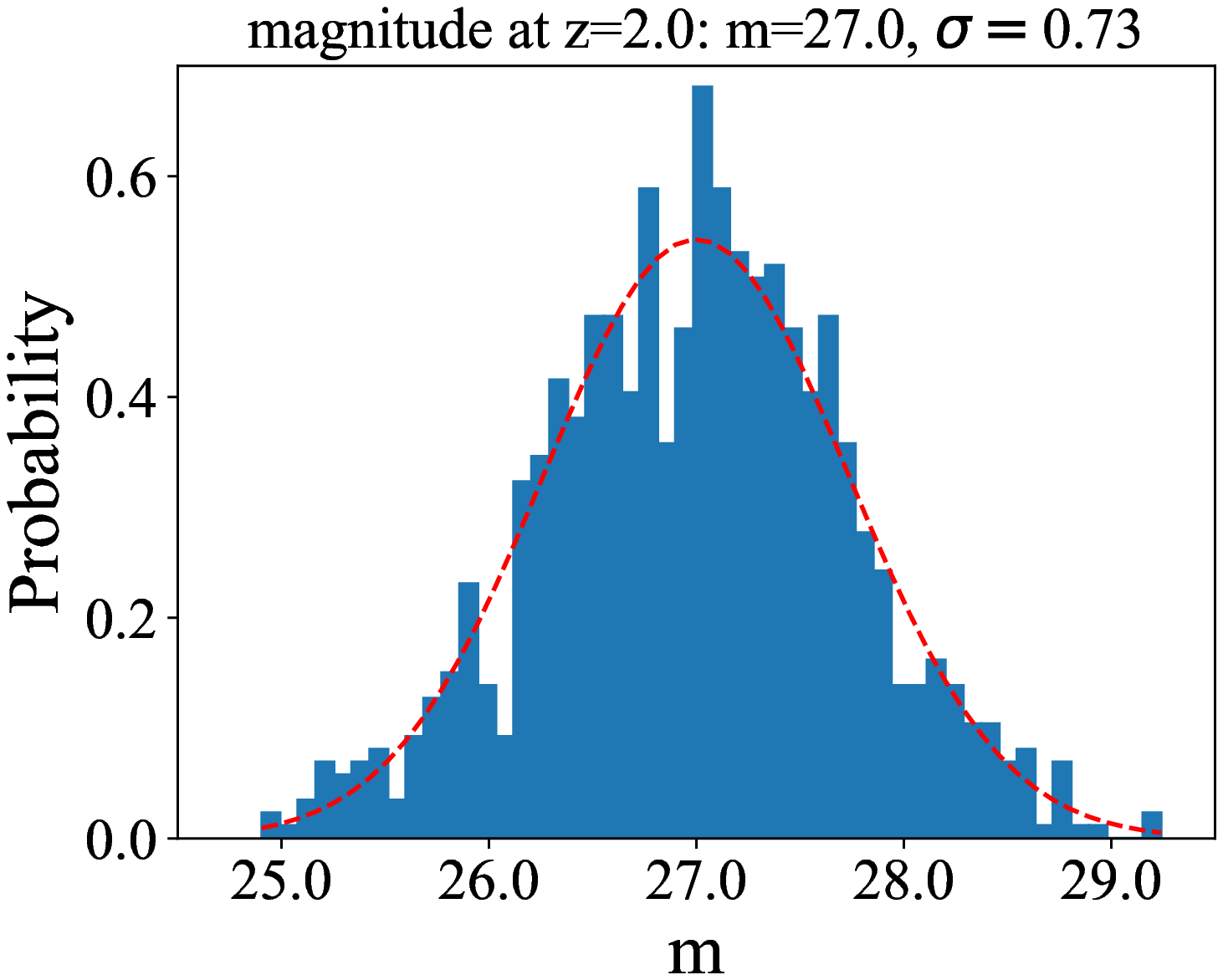}
  \caption{\small{The PDFs of magnitude at redshifts 1.0 (left panel) and 2.0 (right panel), respectively. The red-dashed line is the best-fitting Gaussian distribution.}}\label{fig_hist}
\end{figure}

Now, we start the reconstruction of $m(z)$. Firstly, we normalize the apparent magnitude and sort the data points ($z_i,m_i,\sigma_{m_i}$) in the ascending order of redshift, and re-arrange them into four sequences. In each sequence, the redshifts and the corresponding magnitudes are the input and output vectors, respectively. The inverse of the squared uncertainty ($w_i=1/\sigma_{m_i}^2$) is treated as the weight of data point in the network. Secondly, we train the network constructed as above 1000 times with TensorFlow\footnote{https://www.tensorflow.org} and save this well-trained network. Finally, we execute the trained network 1000 times to predict the magnitude $m$ at any redshift in the range $z\in[0,4]$. Figure \ref{fig_hist} shows the distribution of the magnitude at redshifts $z=1$ and $z=2$ in the 1000 times run of the network. We see that the distribution of magnitude can be well fitted with Gaussian distribution. The mean value and the standard deviation of the Gaussian distribution are regarded as the central value and the $1\sigma$ uncertainty of the prediction, respectively.

\end{multicols}
\begin{multicols}{2}

\begin{figure}[H]
  \centering
  \includegraphics[width=0.8\textwidth]{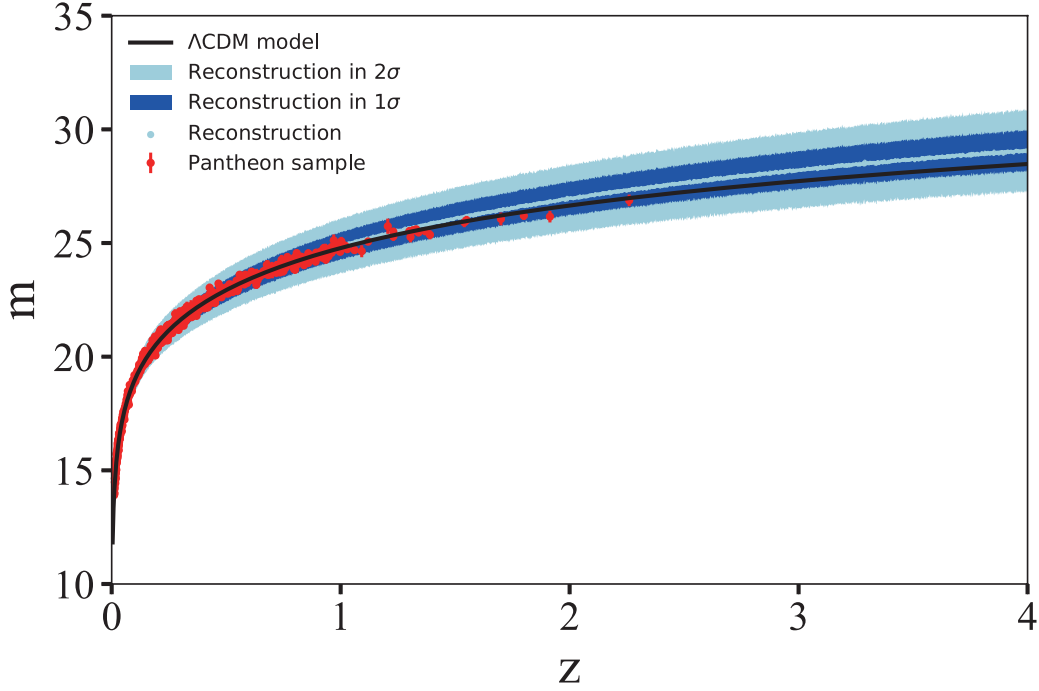}
  \caption{\small{The reconstruction of corrected apparent magnitude-redshift relation $m(z)$ from Pantheon data set. The red dots with $1\sigma$ error bars are the Pantheon data points. The light-blue dots are the central values of the reconstruction. The dark blue and light blue regions are the 1$\sigma$ and 2$\sigma$ uncertainties, respectively. The black line is the best-fitting $\Lambda$CDM curve.}}\label{fig_sim}
\end{figure}

Finally, we obtain the $m(z)$ relation in the redshift range $0<z<4$ and plotted it in Figure \ref{fig_sim}. For comparison, we also plot the best-fitting $\Lambda$CDM curve of the Pantheon sample. We see that the reconstructed curve is well consistent with the $\Lambda$CDM curve, and most of the data points fall into the $1\sigma$ range of the reconstruction. Although the uncertainty of the reconstructed curve using deep learning method is slightly larger than that using GP method in the data region, the merit of deep learning is that it can reconstruct the curve with a relatively small uncertainty beyond the data region, thus allowing us to match SGL with SNe one-by-one, so that the full SGL sample can be used to test DDR.

We note that the uncertainty of the reconstructed $m(z)$ relation is larger than the uncertainty of data points, especially at high redshift. This is because of the sparsity and scattering of data points at high redshift. To check the reliability of the reconstruction, we constrain the matter density parameter $\Omega_m$ of the flat $\Lambda$CDM model using a mock sample generated from the reconstruction, whose redshifts are same as the Pantheon sample, and the magnitudes and the uncertainties are calculated from the reconstructed $m(z)$ relation. Fixing the absolute magnitude $M_B=-19.36$ and Hubble constant $H_0=70$ km/s/Mpc, the matter density is constrained to be $\Omega_m=0.281\pm0.025$, which is well consistent with that constrained from the Pantheon sample, $\Omega_m = 0.278\pm0.008$. This proves that our reconstruction of $m(z)$ relation is reliable.

\end{multicols}
\begin{multicols}{2}

\section{Constraints on DDR}\label{sec:constraints}

\begin{figure}
\centering
\includegraphics[width=0.45\textwidth]{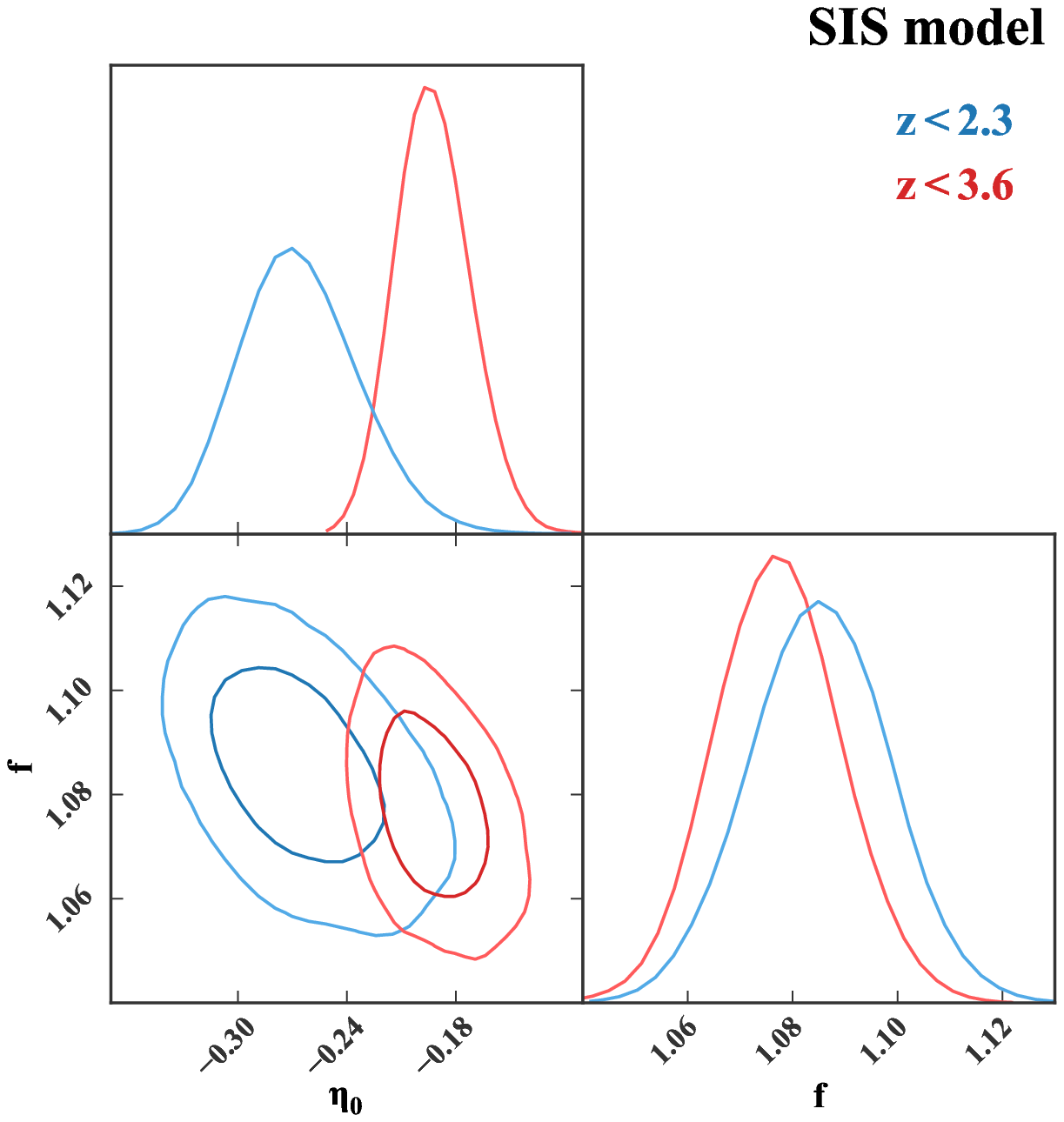}
\includegraphics[width=0.45\textwidth]{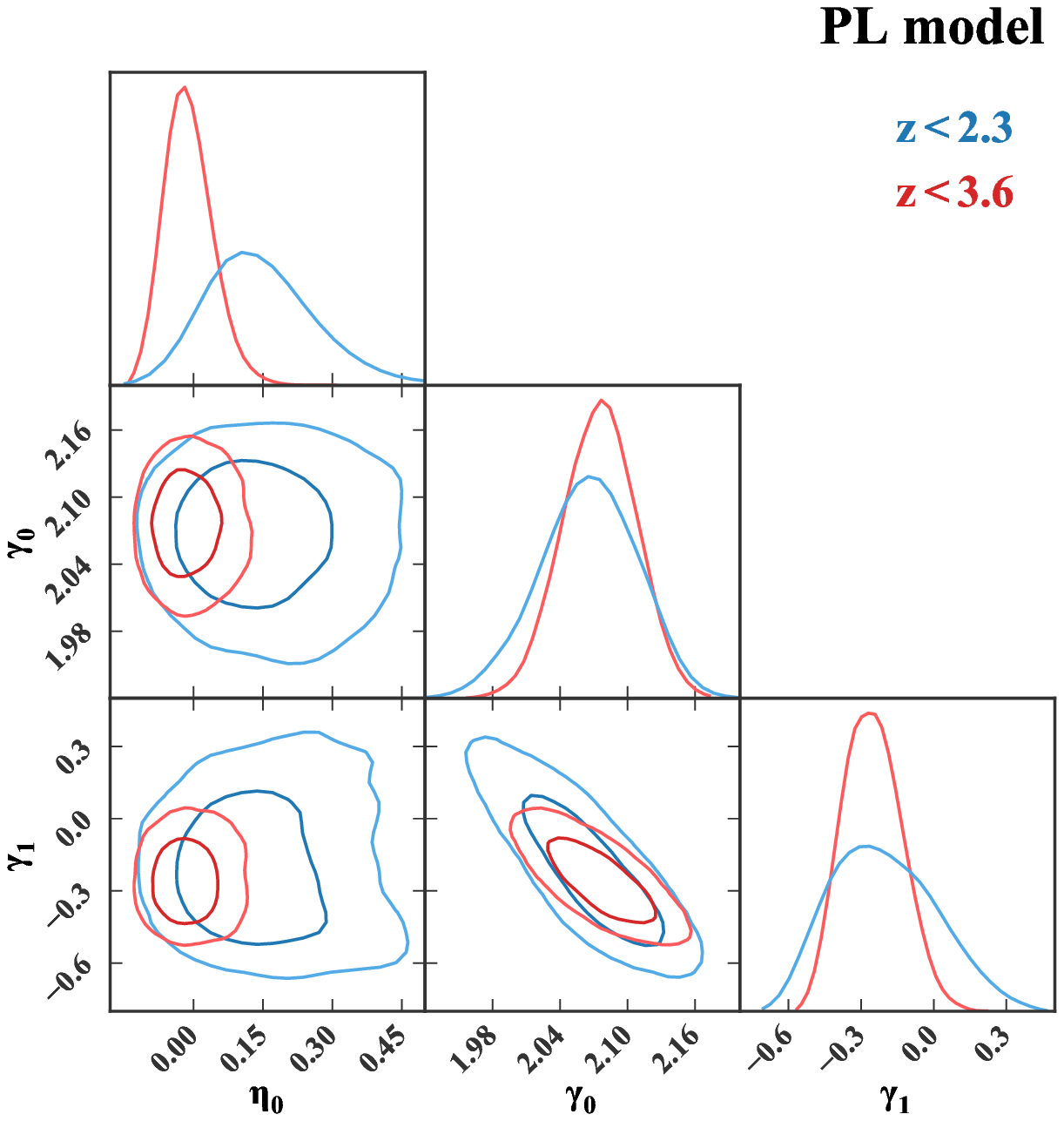}
\includegraphics[width=0.50\textwidth]{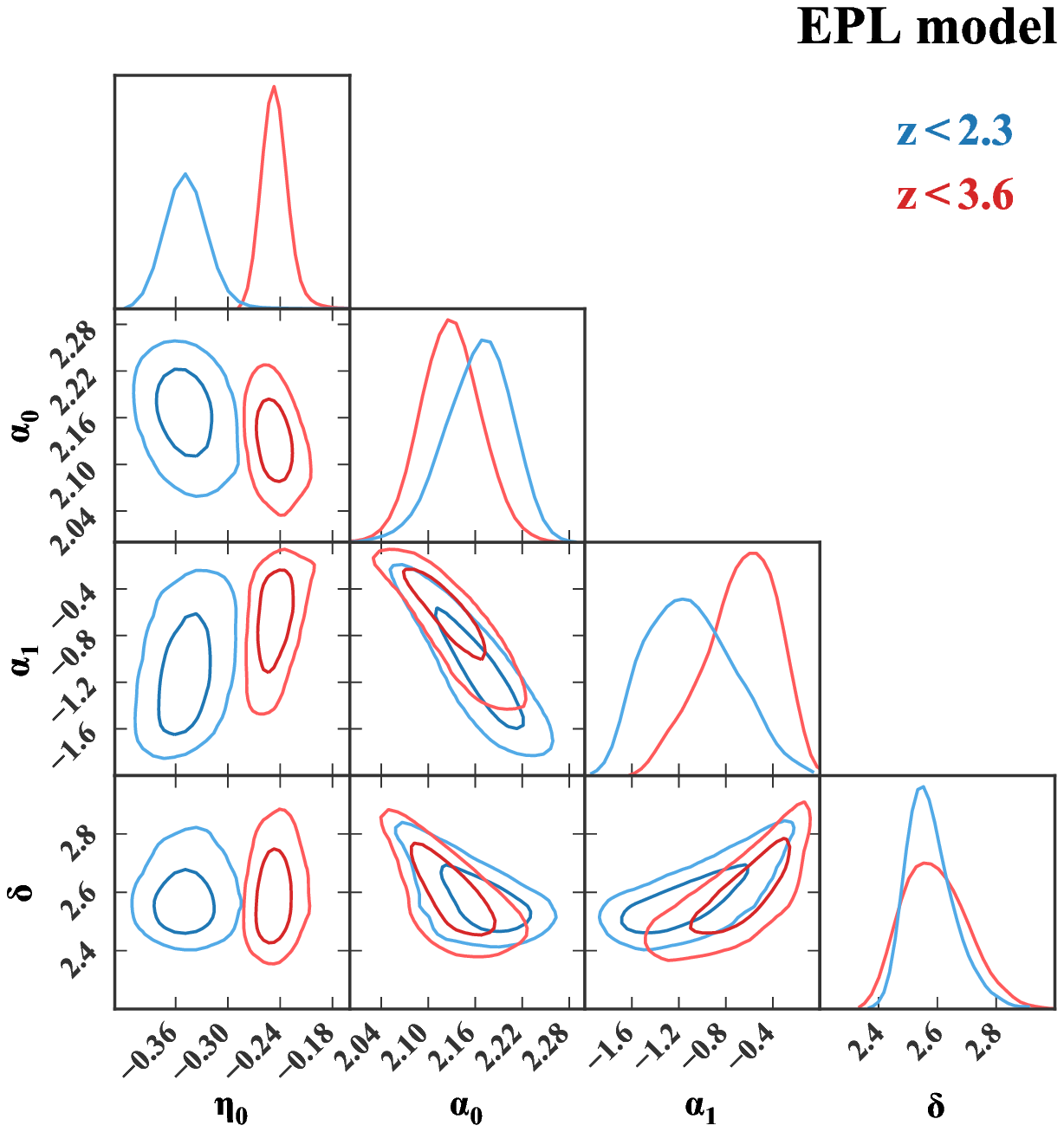}
\caption{\small{The 2-dimensional confidence contours and 1-dimensional PDFs for the parameters in three types of lens models.}}\label{fig:par}
\end{figure}

With the reconstruction of $m(z)$ relation, all of SGL systems are available for testing DDR. To investigate how the inclusion of high-redshift SGL data affects the constraint on DDR, we constrain the parameters with two samples. Sample I includes the SGL data whose source redshift is below $z_s<2.3$, which consists of 135 SGL systems, and sample II includes the full 161 SGL systems in the redshift range $z_s<3.6$. We perform a Markov Chain Monte Carlo (MCMC) analysis to calculate the posterior probability density function (PDF) of parameter space using the publicly available python code \textsf{emcee} \cite{ForemanMackey:2012ig}. Flat priors are used for all free parameters. The best-fitting parameters in SIS, PL and EPL models are presented in Table \ref{tab:parameters}. The corresponding 1$\sigma$ and 2$\sigma$ confidence contours and the posterior PDFs for parameter space are plotted in Figure \ref{fig:par}.

\end{multicols}
\begin{table}[H]
\centering
\caption{\small{The best-fitting parameters in three types of lens models.}}\label{tab:parameters}
\resizebox{!}{0.95cm}
{\begin{tabular}{c|cc|ccc|cccc} 
\hline 
\multicolumn{1}{c|}{}  & \multicolumn{2}{c|}{SIS model}  & \multicolumn{3}{c|}{PL model}  & \multicolumn{4}{c}{EPL model}\\
\hline
Sample     &$\eta_0$     & $f$ &   $\eta_0$     & $\gamma_0$   & $\gamma_1$  &$\eta_0$   & $\alpha_0$   & $\alpha_1$   & $\delta$\\\hline
$z<2.3$ &$-0.268^{+0.033}_{-0.029}$ & $1.085^{+0.013}_{-0.013}$ &
$0.134^{+0.125}_{-0.100}$&	$2.066^{+0.042}_{-0.044}$&	$-0.219^{+0.239}_{-0.205}$ &
$-0.349^{+0.021}_{-0.020}$&	$2.166^{+0.037}_{-0.044}$&	$-1.117^{+0.408}_{-0.358}$&	$2.566^{+0.091}_{-0.066}$\\
$z<3.6$ &$-0.193^{+0.021}_{-0.019}$& $1.077^{+0.012}_{-0.011}$ &
$-0.014^{+0.053}_{-0.045}$&	$2.076^{+0.032}_{-0.032}$&	$-0.257^{+0.125}_{-0.116}$ &
$-0.247^{+0.014}_{-0.013}$&	$2.129^{+0.037}_{-0.037}$&	$-0.642^{+0.274}_{-0.346}$&	$2.586^{+0.117}_{-0.100}$\\
\hline
\end{tabular}}
\end{table}
\begin{multicols}{2}

\end{multicols}
\begin{multicols}{2}

In the framework of SIS model, the DDR violation parameters are constrained to be $\eta_0=-0.268^{+0.033}_{-0.029}$ with sample I and $\eta_0=-0.193^{+0.021}_{-0.019}$ with sample II, which both deviate from zero at $>8\sigma$ confidence level. In the framework of PL model, the constraint of violation parameter is $\eta_0=0.134^{+0.125}_{-0.100}$ with sample I, deviating from the standard DDR at $1\sigma$ confidence level. While the violation parameter is constrained to be $\eta_0=-0.014^{+0.053}_{-0.045}$ with sample II, consistent with zero at $1\sigma$ confidence level. In the EPL model, the violation parameters are $\eta_0=-0.349^{+0.021}_{-0.020}$ with sample I and $\eta_0=-0.247^{+0.014}_{-0.013}$ with sample II, which manifest that DDR is deviated at $>16\sigma$ confidence level. As the results show, the inclusion of high-redshift SGL systems can constrains the DDR violation parameter tighter. Additionally, the model of lens mass profile also has significant impact on the constraint of parameter $\eta_0$. On the premise of exact lens model, DDR can be constrained at a precision of $\textit{O}(10^{-2})$ using deep learning. The accuracy is significantly improved compared with previous results \cite{Liao_2016,Li:2017zrx,Lin:2018qal,Zhou_2021}.

As for the lens mass profile, all parameters are tightly constrained in three lens models. In SIS model, the parameters are constrained to be $f=1.085^{+0.013}_{-0.013}$ with sample I and $f=1.077^{+0.012}_{-0.011}$ with sample II, slightly deviating from unity but with high significance. In PL model, the parameters are constrained with $(\gamma_0,\gamma_1)=(2.066^{+0.042}_{-0.044},-0.219^{+0.239}_{-0.205})$ with sample I and $(\gamma_0,\gamma_1)=(2.076^{+0.032}_{-0.032},-0.257^{+0.125}_{-0.116})$ with sample II. For the slope parameter, it indicates no evidence for the redshift-evolution with sample I. While with sample II, the slope parameter is negatively correlated with redshift at $2\sigma$ confidence level, which is consistent with Chen et al. \cite{Chen:2018jcf}. For the parameters of EPL model, the constraints are $(\alpha_0,\alpha_1,\delta)=(2.166^{+0.037}_{-0.044},-1.117^{+0.408}_{-0.358},2.566^{+0.091}_{-0.066})$ with sample I, and $(\alpha_0,\alpha_1,\delta)=(2.129^{+0.037}_{-0.037},-0.642^{+0.274}_{-0.346},2.586^{+0.117}_{-0.100})$ with sample II. The results demonstrate a non-negligible redshift-evolution of the mass-density slope $\alpha$, which is consistent with Cao et al. \cite{Cao:2016wor}. For all three lens models, none of them can be reduced to the standard SIS model within 1$\sigma$ confidence level. In other word, $f=1$ is excluded in SIS model, $(\gamma_0,\gamma_1)=(2,0)$ is excluded in PL model and $(\alpha_0,\alpha_1,\delta)=(2,0,2)$ is excluded in EPL model.

\section{Discussion and Conclusion}\label{sec:conclusions}

We investigated the distance duality relation with the strong gravitational lensing and SNe Ia using a model-independent deep learning method. With the RNN+BNN network, we reconstructed the apparent magnitude $m$ from the Pantheon compilation up to redshift $z\sim4$. The magnitudes at the redshifts of lens and source in SGL systems can be determined with the reconstructed $m(z)$ relation one-to-one. Compared with previous works \cite{Holanda_2010,Li_2011,Liang2013,Lima:2021slf}, we reconstructed data without any assumption on the cosmological model or the specific parametric form. Compared with GP method \cite{Ruan:2018dls,Wang:2019yob}, our method can reconstruct the data up to higher redshift range but with a lower uncertainty. Taking advantages of all SGL systems and considering the influence of the lens mass profile, we tightly constrained the parameter $\eta_0$ in three lens models. It is found that the constraints on DDR strongly depend on the lens mass model. In the framework of SIS model and EPL model, DDR is deviated at high confidence level. While in the frame work of PL model, not strong evidence for the violation of DDR was found. In other works, if we require that DDR is valid, then the SIS model and EPL model are strongly excluded. The inclusion of high-redshift SGL data does not affect the main conclusions, but can reduce the uncertainty of the DDR violation parameter. In both SIS model and EPL model, the DDR violation parameter $\eta$ favours a negative value, which implies that $D_L>(1+z)^2D_A$. This may be caused by the dust extinction of SNe, making SNe to be fainter (thus has a further luminosity distance) than expected. Once the lens mass model is clear, DDR can be constrained at a precision of $\textit{O}(10^{-2})$ with deep learning, which improves the accuracy by one order of magnitude compared with previous work \cite{Zhou_2021}.

The mass profiles of lens galaxies should be properly considered in cosmological research. We analysed the lens model with the DDR and found that the three types of lens models cannot be reduced to the standard SIS model. With the full SGL sample, the constraining results in SIS model is $f=1.077^{+0.012}_{-0.011}$, which slightly (but with high significance) deviates from the standard SIS model ($f=1$). In PL model, $(\gamma_0,\gamma_1)=(2.076^{+0.032}_{-0.032},-0.257^{+0.125}_{-0.116})$ excludes the standard SIS model $(\gamma_0,\gamma_1)=(2,0)$ at $2\sigma$ confidence level. Similar with the results of Chen et al. \cite{Chen:2018jcf}, the redshift dependence of the slope parameter $\gamma$ in PL model is verified at 2$\sigma$ confidence level. In EPL model, the constraining results are $(\alpha_0,\alpha_1,\delta) =(2.129^{+0.037}_{-0.037}, -0.642^{+0.274}_{-0.346}, 2.586^{+0.117}_{-0.100})$. As can be seen, the total mass profile and the luminosity profile are different due to the influence of dark matter. The slope of mass profile $\alpha$ is obviously redshift-independent, with a trend $\partial \alpha/\partial z_l=-0.642^{+0.274}_{-0.346}$. In order to correctly constrain DDR, accurately modeling the mass profile of the lens galaxies is required.

\end{multicols}

\vspace{-1mm}
\centerline{\rule{80mm}{0.5pt}}
\vspace{2mm}

\begin{multicols}{2}
%

\end{multicols}

\end{CJK*}
\end{document}